\documentclass[11pt,twocolumn]{aastex63}

\received{June 29, 2020}
\revised{July 24, 2020}
\accepted{\today}
\submitjournal{ApJL}

\shorttitle{Fates of clusters after gas expulsion}
\shortauthors{Pang et al.}

\begin{document}

\title{Different Fates of Young Star Clusters After Gas Expulsion}

\author[0000-0003-3389-2263]{Xiaoying Pang}
    \affiliation{Department of Physics, Xi'an Jiaotong-Liverpool University, 111 Ren’ai Road, 
                Dushu Lake Science and Education Innovation District, Suzhou 215123, Jiangsu Province, P. R. China}
    \email{Xiaoying.Pang@xjtlu.edu.cn}
    \affiliation{Shanghai Key Laboratory for Astrophysics, Shanghai Normal University, 
                100 Guilin Road, Shanghai 200234, P. R. China}
    
\author{Yuqian Li}
    \affiliation{Department of Physics, Xi'an Jiaotong-Liverpool University, 111 Ren’ai Road, 
                Dushu Lake Science and Education Innovation District, Suzhou 215123, Jiangsu Province, P. R. China}

\author[0000-0003-4247-1401]{Shih-Yun Tang}
    \affiliation{Lowell Observatory, 1400 W. Mars Hill Road, Flagstaff, AZ 86001, USA}
    \affiliation{Department of Astronomy and Planetary Sciences, Northern Arizona University, Flagstaff, AZ 86011, USA}
        
\author{Mario Pasquato}
    \affiliation{INAF-Osservatorio Astronomico di Padova, Vicolo dell’Osservatorio 5, 35122 Padova, Italy}
    \affiliation{INFN- Sezione di Padova, Via Marzolo 8, I–35131 Padova, Italy}
    
\author[0000-0002-1805-0570]{M.B.N. Kouwenhoven}
    \affiliation{Department of Physics, Xi'an Jiaotong-Liverpool University, 111 Ren’ai Road, 
                Dushu Lake Science and Education Innovation District, Suzhou 215123, Jiangsu Province, P. R. China}

\begin{abstract} 

We identify structures of the young star cluster NGC\,2232 in the solar neighborhood (323.0\,pc), 
and a newly discovered star cluster LP\,2439 (289.1\,pc). Member candidates are identified using 
the {\it Gaia} DR2 sky position, parallax and proper motion data, by an unsupervised machine 
learning method, \textsc{StarGO}. Member contamination from the Galactic disk is further removed 
using the color magnitude diagram. The four identified groups (NGC\,2232, LP\,2439 and two 
filamentary structures) of stars are coeval with an age of 25\,Myr and were likely formed in the
same giant molecular cloud. We correct the distance asymmetry from the parallax error with a 
Bayesian method. The 3D morphology shows the two spherical distributions of clusters NGC\,2232 and
LP\,2439. Two filamentary structures are spatially and kinematically connected to NGC\,2232. 
Both NGC\,2232 and LP\,2439 are expanding.  
The expansion is more significant in LP\,2439, generating a loose spatial distribution with 
shallow volume number and mass density profiles. The expansion is suggested to be mainly driven 
by gas expulsion. NGC\,2232, with 73~percent of the cluster mass bound, is currently experiencing 
a process of re-virialization, However, LP\,2439, with 52 percent cluster mass being unbound, 
may fully dissolve in the near future. The different survivability traces different dynamical 
states of NGC\,2232 and LP\,2439 prior to the onset of gas expulsion. NGC\,2232 may have been 
substructured and subvirial, while LP\,2439 may either have been virial/supervirial, or it has 
experienced a much faster rate of gas removal.

\end{abstract}

\keywords{stars: evolution --- open clusters and associations: individual (NGC\,2232, LP\,2439) -- stars: kinematics and dynamics}

\section{Introduction}\label{sec:intro}

Star clusters form from dense gas clumps inside giant molecular clouds (GMCs). 
Only a fraction of the gas in a GMC is converted into stars, while the remainder of the gas 
is expelled from the cluster-forming clumps at later times. Therefore, the dynamics of star clusters embedded in the molecular 
cloud is dominated by the gravitational potential of the gas from which star clusters 
are formed \citep{lada2003}. Feedback from high-mass stars, such as supernovae, stellar winds,
and radiation, will remove the intra-cluster gas within several million years, which 
rapidly reduces the gravitational potential of the star cluster. This results in a star cluster 
expansion over a period of $4-10$~Myr, from an initial radius of $1-2$~pc, depending on how rapidly 
the gas is removed \citep[see][]{kroupa2005, baumgardt2007}. Most star clusters will lose a 
significant fraction of their stellar mass after the phase gas dispersal, and often completely 
dissolve into the galactic field. This is usually referred to as infant mortality \citep{deGrijs2008}. 
In recent years, many studies investigated the processes and effects of gas dispersal on the 
evolution of young clusters \citep[e.g.,][and references therein]{hill1980, goodwin2006, baumgardt2007,
banerjee2013, brinkmann2017, Shukirgaliyev2017, farias2018, dinnbier2020a, dinnbier2020b}. 

The dynamical state of the star cluster at the onset of gas expulsion plays a crucial role in 
determining the the survival chances of a star cluster after the gas has been removed. If the gas 
expulsion occurs when the cluster is dynamically subvirial and has a substructured spatial 
distribution, the survival rate of the cluster is considerably higher 
\citep{goodwin2009, farias2015, farias2018}. Moreover, the faster the gas is expelled from the cluster,
the lower the bound fraction of mass for the remaining cluster members is \citep{brinkmann2017}. 
If the stellar mass of a protocluster is more centrally-concentrated than that of the gas, as 
obtained in the cluster formation model of \citet{parmentier2013}, the survival rate of star clusters is 
also much higher \citep{adams2000,Shukirgaliyev2017}. 

The young star clusters that emerge from their parental molecular clouds shortly after gas
expulsion will enter a phase of violent relaxation, which is the phase in which the star clusters 
evolve from a non-equilibrium state towards a new state of equilibrium \citep{lynden1967}. 
If the cluster succeeds in obtaining virial equilibrium during this phase, it is said to have been re-virialized. 
Several young massive clusters, such as R\,136, and Westerlund\,1, manage to 
regain virialization within a time shorter than their present age \citep{banerjee2013, cottaar2012}. 
On the other hand, the re-virialization time is substantially longer than the age of several other 
young clusters, for example in the cases of IC\,2602, IC\,2391, and NGC\,2547 \citep{bravi2018}. 
These clusters are supervirial and continue to expand.

In addition to gas expulsion, the two-body relaxation process in a cluster can also trigger 
expansion of the cluster. However, the expansion rate induced by two-body relaxation is 
substantially smaller than that resulting from gas expulsion 
\citep[see, e.g.][]{kroupa2005, moeckel2012, dinnbier2020a, dinnbier2020b}.  Simultaneously, mass segregation 
occurs as a byproduct of two-body relaxation, as the cluster members attempt to achieve a state of energy-equipartition \citep[e.g.,][]{pang2013}. Thus far, no astrometric efforts have 
been made to distinguish these two drivers of expansion in star clusters. Testing these scenarios using observations of young star clusters shortly after gas expulsion has occurred, is crucial to understand the disruption process of star clusters.

NGC\,2232 is a young open cluster with an age of $25-29$\,Myr \citep{currie2008, Liu2019}, 
located south of the Galactic plane at the position of R.A.=$96\fdg9973$, Decl.=$-04\fdg7929$
\citep{gaia2018b}. Unlike its neighbor, the Orion star-forming complex, NGC\,2232 has not 
attracted much attention. There exist only several comprehensive studies of this cluster. 
\citet{claria1972} carried out the first $UBV$-photometry study of this cluster and estimate 
its age to be 20\,Myr. They measured the interstellar reddening of the cluster as $E(B-V)=0.01$\,mag,
and obtained a distance of 360\,pc. The latter measurement of the reddening was refined to $0.06\pm0.03$\,mag 
in the first spectroscopic study of sixteen cluster members by \citet{levato1974}. \citet{lyra2006} constrained 
the reddening in the colour–colour diagram by studying eight evolved B-type stars, which gave a 
best-fitting reddening of $E(B-V) = 0.07\pm0.02$\,mag. 

Young star clusters that are still partially embedded in the gas and dust of their parental molecular
clouds tend to have a high interstellar reddening. This phenomenon is seen in the Orion Nebula Cluster, which has 
differential reddening of several orders of magnitude  \citep{Scandariato2001}.
Also, the reddening of the 1\,Myr old cluster NGC\,3603 can reach 1.5\,mag \citep{pang2011}. The low reddening 
of NGC\,2232 thus indicates that it probably has already undergone gas expulsion.

Located right next to NGC\,2232 (within 50\,pc) is a newly-discovered star cluster,  LP\,2439, that
was recently found by \citet[ID:2439]{Liu2019}. This star cluster has not appeared in any study since 
its discovery in 2019. According to \citet{Liu2019}, LP\,2439 has an age similar to that of NGC\,2232. 
Note that a common age does not necessarily imply a similar dynamical evolutionary stage, as the dynamical ages of star clusters, which may be quantified using the number of crossing times since their formation (see Section~\ref{sec:discussion}).
We therefore select both clusters as targets to investigate the expansion and dynamical state of star 
clusters after gas expulsion.

Using the data from {\it Gaia} DR\,2, we explore the neighborhood of NGC\,2232 and LP\,2439, and 
investigate their dynamical state after gas expulsion. In Section~\ref{sec:gaia_member}, we discuss
the quality and limitations of the {\it Gaia} DR\,2 data, and describe our input data-set for structure
identification. We then present the algorithm, \textsc{StarGO}, which is used to identify structures.
Further confirmation of membership is performed using the the color-magnitude diagram. The properties 
of the identified member candidates are presented and discussed in Section~\ref{sec:result}, including 
the discussion of the evidence of gas expulsion. The 3D morphology of NGC\,2232, LP\,2439, and
the nearby associated structures are discussed in Section~\ref{sec:3D} along with our interpretation of
the results. The fate and dynamical state of both clusters are discussed in Section~\ref{sec:discussion}.
Finally, we provide a brief summary in Section~\ref{sec:summary}.

\section{Data Analysis and Member Determination}\label{sec:gaia_member}

\begin{figure*}[tb!]
\centering
\includegraphics[angle=0, width=1.\textwidth]{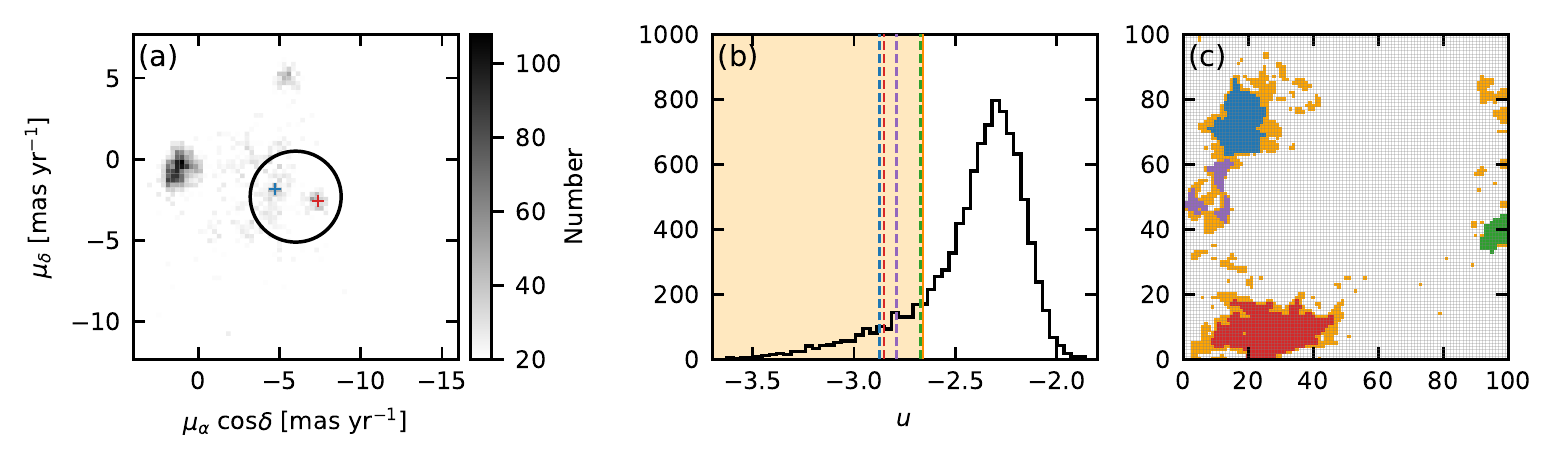}
\caption{ 	 
    (a)~ the 2D density map of the proper motion vectors for the regions around NGC\,2232 and
         LP\,2439 in Sample~I. The blue and red crosses indicate the over-densities generated 
         by NGC\,2232 and LP\,2439, respectively. Each bin is smoothed by neighboring 8 bins 
         and here only bins with a number count $>3\sigma$ are shown, where $\sigma$ is the 
         standard deviation of all bins. The color bar indicates the number count in each bin.
	(b)~ Histogram of the distribution of $u$. The orange solid line shows the position 
	     $u_{\mathrm{peak}-3\sigma}$. The left tail with $u<u_{\mathrm{peak}-3\sigma}$ is
	     highlighted in orange, corresponding to the orange patches in panel (c).
	     The blue, red, purple, and green dashed lines denote the threshold values of $u$. 
	     Values smaller than this threshold produce the 5\% contamination rate among identified candidates for 
	     the red, blue, purple and green patches in SOM (panel (c)). 
	(c)~ 2D 100 $\times$ 100 neural map resulting from SOM, the neurons with $u$ 
	     smaller than the threshold values (blue, red, purple and green dashed lines in panel (b))
	     of the 5\% contamination rate are colored as blue, red, purple and green. The two dominant neuron 
	     groups are traced with blue (NGC\,2232) and red (LP\,2439). 
	    }
\label{fig:som}
\end{figure*}

\subsection{Gaia DR\,2 Data Processing and Analysis}\label{sec:target_selection}

The second data release (DR2) of {\it Gaia} \citep{gaia2018a} has provided more than one billion 
sources with parallaxes ($\varpi$) and proper motions (PMs; $\mu_\alpha \cos\delta, \mu_\delta$) 
with unprecedented precision and sensitivity. The $G$~band ($330-1050$~nm) photometry ranges from 
$\sim 3$~mag to 21~mag. The median uncertainty of $\varpi$ ranges from $\sim$0.04~mas for bright 
sources ($G$\,$<$\,14~mag) to $\sim$0.7~mas for faint ones ($G$\,$\approx$\,20~mag). The corresponding 
uncertainties of PMs for these sources are 0.05\,mas\,yr$^{-1}$ and 1.2\,mas\,yr$^{-1}$, respectively 
\citep{lindegren2018}. Only 7.2~million stars have radial velocity (RV) measurements in the {\it Gaia} 
DR\,2 \citep{cropper2018} with a typical uncertainty of 2~$\rm km\,s^{-1}$. 

The spatial and kinematic structures of NGC\,2232 and its neighboring cluster LP\,2439 are investigated 
using {\it Gaia} DR\,2 data within 100\,pc from the center of NGC\,2232. In Cartesian Galactocentric 
coordinates, the center of NGC\,2232 is located at ($X,Y,Z$) = ($-8566.0$, $-183.1$, $-14.0$)~pc 
\footnote{
The Cartesian Galactic coordinate system adopted in our study is defined as follows. The Galactic center ($l=0\degr$ and $b=0\degr$) is located at the origin of the coordinate system. The positive $X$-axis points from the projection of the Sun's position onto the Galactic mid-plane towards the Galactic center.
The positive $Y$-axis points towards $l=90\degr$, and the positive $Z$--axis points 
towards $b=90\degr$. Further details about the coordinate system can be found in the appendix of \citet{tang2019}.
}. 
The distance to NGC\,2232 is taken to be 325.6~pc \citep{gaia2018b}, and the equatorial coordinates of its center as (R.A.=$96\fdg9973$, Decl.=$-04\fdg7929$, J2000) from \citet{gaia2018b}. 

We apply the same astrometric quality cuts as \citet[in their Appendix C]{lindegren2018} to exclude 
possible artifacts in the {\it Gaia} DR\,2 from our sample. The final ``cleaned'' sample contains 131,066
sources. Hereafter, we refer to this set as ``Sample~I''. The stars in sample~I have $G$ 
magnitudes ranging between $\sim$4.2~mag and $\sim$19.5~mag, and the sample becomes significantly incomplete for 
$G \ga 19$~mag.

Figure~\ref{fig:som}~(a) shows a 2D density map of PMs that only shows bins with over-densities 
$>3\sigma$ in Sample~I. Several over-densities stand out. There is an over-density near
the average PMs of NGC\,2232 (indicated with a blue cross)  
($\mu_\alpha \cos\delta$, $\mu_\delta$)=($-4.734$, $-1.844$)~$\rm mas~yr^{-1}$
provided by \citet{Liu2019}. Another over-density is seen at 
($\mu_\alpha \cos\delta$, $\mu_\delta$)=($-7.381$, $-2.569$)~$\rm mas~yr^{-1}$ 
\citep[red cross,][]{Liu2019}. 
This group corresponds to LP\,2439.The notable clustering at $\mu_\alpha \cos\delta<0$ $\rm mas~yr^{-1}$ and
$\mu_\delta<0$ $\rm mas~yr^{-1}$ is the Orion star forming complex \citep{jerabkova2019}. An isolated over-density 
around ($\mu_\alpha \cos\delta$, $\mu_\delta$)=($-5.310$, $5.014$)~$\rm mas~yr^{-1}$ relates to 
another new star cluster discovered by \citet{Liu2019} (ID: 2383), which is located at a distance of 377\,pc \citep{Liu2019} and does not show any spatial connection to 
NGC\,2232. We apply a circular cut with a radius of 2.8~$\rm mas~yr^{-1}$ (the black circle in
Figure~\ref{fig:som}~(a)), to only include NGC\,2232 and LP\,2439, for further analysis. 
This reduces the number of stars to 5,843; we refer to this set of stars as "Sample~II", hereafter.
The stars in this sample have magnitudes ranging between $G \sim 5.2$~mag and $G \sim 19.3$~mag, and the sample is complete 
for $G\la 18.0$~mag. Note that the stars in Sample~II are typically fainter than the stars in 
Sample~I, since the brightest stars, which reside in the Orion star forming complex, have been 
removed.

In this study we use 5D parameters of stars in Sample~II 
(R.A., Decl., $\varpi$, $\mu_\alpha \cos\delta$, and $\mu_\delta$) from {\it Gaia} DR\,2. Since 
only a fraction of the stars ($\sim$12 percent) have RV measurements, RV measurements are used as 
supplementary data. Adopting a distance of $1/\varpi$, we compute for each source the Galactocentric 
Cartesian coordinates ($X, Y, Z$). The transformation is performed by using the Python \texttt{Astropy} 
package \citep{astropy2013, astropy2018}. Considering the asymmetric error in the distance that arises from the direct 
inversion of $\varpi$, a Bayesian method is adopted to correct individual distances of stars in 
Section~\ref{sec:3D}.

\subsection{Member identification}\label{sec:stargo}

The unsupervised machine learning method,
\textsc{StarGO} \citep{yuan2018}\footnote{\url{https://github.com/salamander14/StarGO}} is 
used for member candidate selection. This algorithm is based on the Self-Organizing-Map (SOM) 
method that map high-dimension data down to two dimensions, while preserving the topological 
structures of the data. \textsc{StarGO} has been successfully used in the identification of 
tidal structures in open clusters, such as the Coma Berenices cluster \citep{tang2019} and the 
Blanco\,1 \citep{zhang2020}.

We apply \textsc{StarGO} to map a 5D data set ($X, Y, Z$, $\mu_\alpha \cos\delta, \mu_\delta$) 
of the clusters NGC\,2232, LP\,2439 and their surrounding region (Sample~II) onto a 2D neural network to 
identify member candidates. We adopt a network with 100$\times$100 neurons represented 
by the 100$\times$100 grid elements in Figure~\ref{fig:som}~(c), to study Sample~II. 
Each neuron is assigned with a random 5D weight vector that has the same dimensions as the input 5D parameters ($X, Y, Z$, $\mu_\alpha \cos\delta, \mu_\delta$) obtained from observations. The weight vector of each neuron is updated during each iteration so that it is closer to the input vector of an observed star. This learning process
is iterated 400 times until the weight vectors converge. The value of the difference of weight vectors between 
adjacent neurons, $u$, is small when the 5D weight vectors of the adjacent neurons are similar, indicating stars
associated with neurons are spatially 
and kinematically coherent. The neurons with small values of $u$ will group together in 
the 2D SOM map as patches (see Figure~\ref{fig:som}~(c)). Different patches correspond to
different groupings of stars. Neurons located inside the patch have smaller $u$ value than neurons 
outside. Therefore, $u$ values of neurons inside patches generate an extended tail on the left 
(toward small value) in the $u$ histogram (see Figure~\ref{fig:som}~(b)).

A cut on the tail of the $u$ distribution is used for member selection. \citet{tang2019} and \citet{zhang2020} 
select member candidates in the extended tail of the $u$ distribution, with $u$ values less than the 
value of the peak $u_{\mathrm{peak}}$ minus the 99.85 percentile 
$\Delta_{3\sigma}$ ($u\le u_{\mathrm{peak}}-\Delta_{3\sigma}$; see their figure~3). Note that our 
samples cover the Galactic mid-plane. Therefore, contamination by field stars with similar PMs to 
those of cluster members is much higher than for Coma Berenice \citep{tang2019} and Blanco\,1 \citep{zhang2020, jackson2020},
which are $100-200$\,pc off the mid-plane. Applying the $u$-cut criteria, 
$u\le u_{\mathrm{peak}}-\Delta_{3\sigma}$ (orange vertical line in Figure~\ref{fig:som} (b)) results 
in a 13.3 percent field star contamination rate in selected patches showed up in the SOM (oragne
patches in Figure~\ref{fig:som} (c)), which is double the value of $\sim5-6\%$ in Coma Berenice 
\citep{tang2019} and Blanco\,1 \citep{zhang2020}. 

The contamination rate is evaluated from the smooth Galactic disk population using the {\it Gaia} 
DR\,2 mock catalog \citep{rybizki2018}, by applying the same PM cut as described in 
Section~\ref{sec:target_selection} to the mock catalog in the same volume of the sky. Each of 
these mock stars is attached to the trained SOM map, from which we can compute the number of field 
stars associated with selected patches. 
 
In order to reduce field star contamination, the selection of $u$ is chosen to ensure a similar contamination 
rate of $\sim$5\% for the different selected patches (blue, red, purple, and 
green dashed lines in Figure~\ref{fig:som} (b)). We color the two major clusters, NGC\,2232 and
LP\,2439 in blue and red, respectively. The two additional smaller groups are colored as purple and green. Three small purple 
patches merge into NGC\,2232 as one big group (orange patches) if we apply a $u$-cut of
$u_{\mathrm{peak}}-\Delta_{3\sigma}$. Considering their similarity, purple patches are considered as 
one single group. Properties of NGC\,2232, LP\,2439, and the neighboring groups will be discussed 
further in Section~\ref{sec:result}. In the end, we have 182 member candidates for NGC\,2232 (blue), 
331 for LP\,2439 (red), 79 for the purple group, and 61 for the green group.

\subsection{Member Cleaning via the Color Magnitude Diagram}\label{sec:cmd_clean}

\begin{figure*}[tb!]
\centering
\includegraphics[angle=0, width=0.95\textwidth]{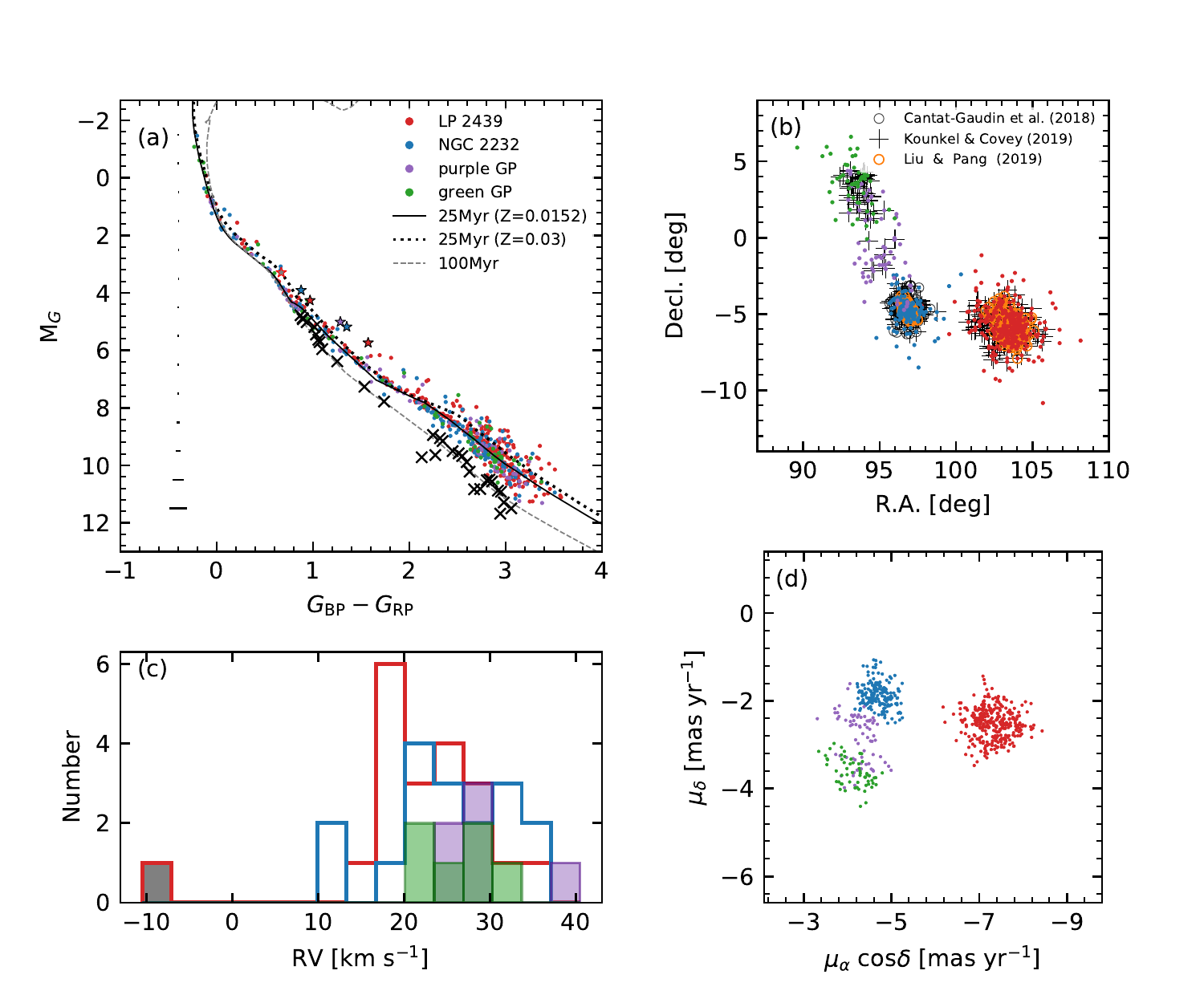}
\caption{ General properties of identified member candidates. 
        (a)~ The color magnitude diagram of {\it Gaia} DR\,2 absolute magnitude M$_{G}$ 
             (adopting {\it Gaia} DR\,2 parallax) for candidate stars identified by \textsc{StarGO}.
             Red and blue dots are candidate members of LP\,2439 and NGC\,2232. The purple and green dots 
             indicate candidates of the purple and green groups in Figure~\ref{fig:som}~(c). Crosses 
             indicate field contaminants that are excluded from further investigation. The PARSEC 
             isochrones of 25\,Myr and 100\,Myr are overplotted as the black solid curve and grey 
             dashed curve, with $E(B-V)=0.07$~mag \citep{lyra2006} corrected and solar metallicity 
             assumed. The black dotted curve is the PARSEC isochrone of 25\,Myr adopting the same reddening 
             but a supersolar metallicity of [Fe/H]=0.32 \citep{monroe2010}. The asterisks indicate 
             binary candidates. The red outlined grey shaded asterisk corresponds to the grey shaded bin 
             in panel (c). Uncertainties of colors are indicated with horizontal error bars. The typical 
              $G$ magnitude error is 0.002\,mag.
        (b)~ The spatial distribution of member candidates after CMD cleaning. The symbols and colors of
             member candidates are identical to those in panel (a). Members cross-matched with previous catalogs:
             \citet{cantat2018}, black open circles; 
             \citet{kounkel2019}, black crosses; 
             \citet{Liu2019}, orange open circles. 
        (c)~ is the RV histograms for LP\,2439 (red), NGC\,2232 (blue), the purple group and the green group. 
        (d)~ shows the proper motion vector plot with colors and symbols the same as panel (a) for 
             member candidates. 
	    }
\label{fig:properties}
\end{figure*}

We construct a color-magnitude diagram (CMD) for the 653 candidate stars of the four groups in
Figure~\ref{fig:properties}~(a). Stars in the four groups surprisingly track a clear locus of 
a main-sequence together, which is consistent with the PARSEC isochrone of 25\,Myr with the 
sensitivity curves provided by \citet{wei18} (black solid curve), adopting a reddening 
$E(B-V)=0.07$\,mag \citep{lyra2006} and a solar metallicity. Based on the nice fit of the isochrone 
to the data, the adopted age is in agreement with \citet{currie2008}, who obtained the age of 
NGC\,2232 using ROSAT x-ray members. Different groups being coeval implies a common origin (see 
discussion in Section~\ref{sec:result}). Although NGC\,2232 is suggested to be a metal-rich 
open cluster with a metallicity of [Fe/H]=$0.32 \pm 0.08$ \citep{monroe2010}, the isochrone with 
super-solar abundance (dotted curve in Figure~\ref{fig:properties}~(a)) is clearly shifted
from the main sequence locus towards redder colors due to the higher metallicity. Note that 
\citet{monroe2010}'s measurement was based on spectroscopic observations of four pre-main-sequence stars, 
whose lithium abundance is higher than that of the main-sequence stars due to lower effective temperatures
\citep{juarez2014}. 
  
A dozen of candidate stars are located below the main sequence locus ($M_G>4.7$\,mag), corresponding 
to an age equal to or older than 100\,Myr 
(grey dashed curve in Figure~\ref{fig:properties}~(a)). These stars are located on the outskirts of NGC\,2232
and LP\,2439. The RV of one star among these is four times larger than that of the other member candidates. 
Therefore, these are probably field stars. We use the 100\,Myr isochrone as a reference
and redden it to exclude possible contaminants. Around M$_G=8$\,mag is the transition region between
pre-main-sequence stars to main-sequence stars. We redden the 100\,Myr isochrone by 0.04\,mag for 
magnitudes brighter than M$_G$=8\,mag, and by 0.1\,mag for those with M$_G>$8\,mag. Stars bluer than the reddened 
isochrone are excluded from further analysis (black crosses in Figure~\ref{fig:properties}~(a)). After 
the CMD cleaning, 621 stars are selected as member candidates, 
NGC\,2232 (blue): 177, LP\,2439 (red): 315, purple: 71; green: 58.

\section{Properties of Identified Groups}\label{sec:result}

\subsection{Crossmatch with Previous Catalogs}\label{sec:match}

Several new catalogs have recently been published for identifying new stellar groups with
the multi-dimensional parameters from {\it Gaia} DR\,2. Recent work by \citet{Liu2019} has 
identified star clusters in {\it Gaia} DR\,2 using the friend-of-friend (FoF) cluster finder in 
the five-dimensional parameter space  ($l, b, \varpi, \mu_\alpha\cos\delta$, and $\mu_\delta$). 
LP\,2439 (the red group), is one of the 76 new clusters they discovered. Among their 148 members 
for LP\,2439, we recover 129 (orange circles in Figure~\ref{fig:properties}~(b)) in the present
study. In addition, among the 177 member candidates that we identify in NGC\,2232, 91 are 
cross-matched with the membership list of LP\,2393 in \citet[][ID:2393]{Liu2019}. 
As can be seen in Figure~\ref{fig:properties} (b), members identified with the FoF cluster finder are all 
concentrated near the centers of the two clusters. \citet{cantat2018} developed an unsupervised membership 
assignment code \textsc{UPMASK} to identify star clusters. In \citet{cantat2018}'s study,  
151 of the member candidates in NGC\,2232 (blue) and 6 of the member candidates in the purple group (black circles in 
Figure~\ref{fig:properties}~(b)) are identified as members of NGC\,2232 in \citet{cantat2018}.

\citet{kounkel2019} applied an unsupervised machine learning techniques and identified 1900 star 
clusters and strings in the Galactic disk. The strings that they identified are co-moving groups
with filamentary structures reaching up to $\sim$200\,pc in length. They identified 328 strings 
with ages younger than 100\,Myr in the Solar neighborhood. \citet{kounkel2019} classify NGC\,2232 
as a string spanning 40 degrees in the sky. The membership list of this NGC\,2232 string in \citet{kounkel2019} 
cross-matches with 236 member candidates in LP\,2439, 142 in NGC\,2232, 32 candidates in the purple group, 
and 30 candidates in the green group (black crosses in Figure~\ref{fig:properties}~(b)). 
Stars in strings are not only kinematically and spatially coherent, but also coeval. \citet{kounkel2019}
suggest that the extended shape of strings are primordial structures that originated from giant molecular 
filaments \citep{zucker2018}. 

Unlike \citet{kounkel2019}'s claim, of a ``lack of a central cluster" among strings, NGC\,2232 and LP\,2439 
do not only stand out as two central clusters with significant overdensities in the space, but also 
show two distinct overdensities in the PM distribution (Figure~\ref{fig:properties}~(d)). Their RV 
distributions slightly offset each other (Figure~\ref{fig:properties}~(c)), with an average RV of 
22.0~km~s$^{-1}$ and 25.4~km~s$^{-1}$ for LP\,2439 and NGC\,2232, respectively. The mean RV value obtained 
from {\it Gaia} DR\,2 for NGC\,2232 agrees well with
the mean RV of 25.40~km~s$^{-1}$ that was measured by 
\citet{jackson2020} using the {\it Gaia}-ESO Survey. Considering the coevality of the four groups, 
they were likely formed during the same star formation event in the same GMC. Two local high-density  
regions in the GMC formed clusters, NGC\,2232 and LP\,2439. The purple and green groups consist of stars 
formed along the primordial filaments associated with NGC\,2232, since their PM distributions are similar to
NGC\,2232, instead of LP\,2439 (Figure~\ref{fig:properties}~(b)).

There is one member candidate in LP\,2439 with a negative RV, $-$9.73~$\rm km~s^{-1}$ (the grey shaded bin 
in Figure~\ref{fig:properties}~(c)), different from the majority in LP\,2439. Considering its location along 
the binary sequence (the red outlined grey shaded asterisk in Figure~\ref{fig:properties}~(a)), we conclude that
it is a probable binary candidate. The other five member candidates with RV measurements are also binary 
candidates, being located along the binary sequence (filled asterisks in Figure~\ref{fig:properties}~(a)). 
Further follow-up spectroscopy may further constrain the nature of these stars. 

\subsection{Evidence of Gas Expulsion}\label{sec:match}

\begin{figure}[tb!]
\centering
\includegraphics[angle=0, width=1.\columnwidth]{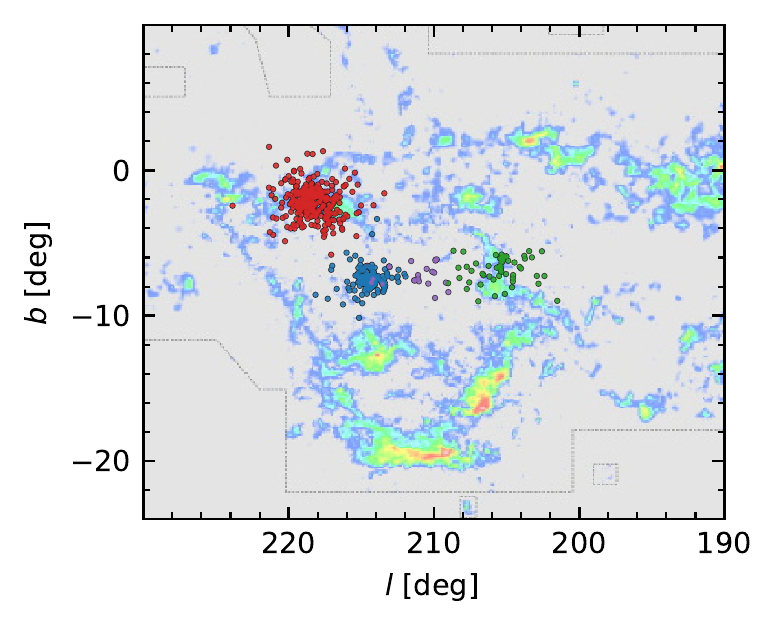}
\caption{
        CO emission map obtained from \citet{dame2001} in the region where member candidates reside. 
        The red dots are candidate members of LP\,2439, and the blue dots are candidate members of 
        NGC\,2232. The purple and green dots represent the purple and green groups, respectively.
	    }
\label{fig:CO}
\end{figure}

As indicated by the low reddening in NGC\,2232 \citep{lyra2006}, the gas content is very low in the 
region occupied by our member candidates. We over-plot member candidates on the map of a large scale 
CO survey of the Galactic plane \citep[][Figure~\ref{fig:CO}]{dame2001}. Towards the south of NGC\,2232, 
are remarkably long and thin molecular filaments at $ 200\degr \la l \la 220\degr$ and
$-20\degr \la b \la -10\degr$, which belong to the Orion cloud in the background 
\citep[No. 27 in Table 1 in ][Figure~\ref{fig:CO}]{dame2001} located at a distance of $414-437$\,pc 
\citep{hirota2007, menten2007}. LP\,2439 happens to overlap with the CO emission at 
$-3\degr \la b \la 0\degr$ 
\citep[No. 31 in Table 1 in ][Figure~\ref{fig:CO}]{dame2001}, which is 
induced by molecular clouds at a distance of 12--18\,kpc from the Galactic center \citep{may1993}. 
The Leiden-Dwingeloo 21-cm survey \citep{hartmann1997} also shows a very low column density of atomic 
hydrogen in the region of $200\degr \la l \la 220\degr$ and $-10\degr \la b \la 0\degr$.
Therefore, the region at the distance of LP\,2439, NGC\,2232 and its related filamentary structures 
(the purple and green groups) can be considered free of gas and dust. This is direct evidence that the 
parental molecular gas from which the stars in NGC\,2232 and LP\,2439 were born, has already been 
expelled by supernovae and/or stellar winds. The two clusters enter a phase of violent relaxation after
gas expulsion (see discussion in Section~\ref{sec:discussion}).

\section{3D Morphology of NGC\,2232 and LP\,2439}\label{sec:3D}

\subsection{Distance Correction}\label{sec:dis_correct}

\begin{figure*}[tb!]
\centering
\includegraphics[angle=0, width=.95\textwidth]{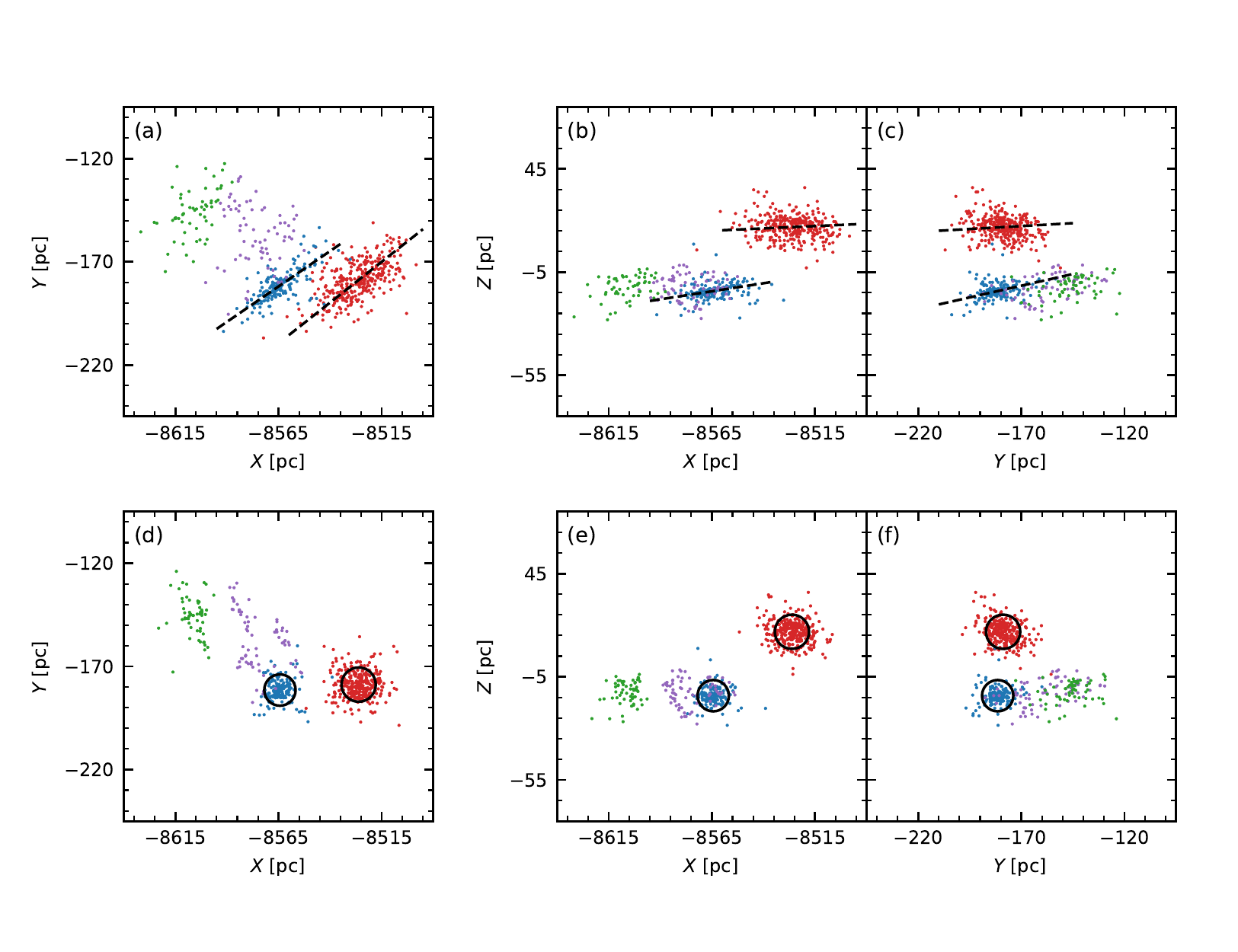}
\caption{
        3D spatial position of member candidates in Galactocentric Cartesian $X,Y,Z$ coordinates. 
        The colored symbols are the same as Figure~\ref{fig:properties}~(a). 
        (a)--(c)~display positions before distance correction. The black dashed lines indicate the
                direction of the line of sight. 
        (d)--(f)~show 3D positions after distance correction via a Bayesian approach 
        (see Section~\ref{sec:dis_correct}). 
	    }
\label{fig:3d}
\end{figure*}

In Figure~\ref{fig:3d}, panels (a) to (c), we show the 3D spatial distributions of member 
candidates in NGC\,2232, in LP\,2439, in the purple group, and in the green group. The shapes 
of these four groups are all stretched along the same direction: the line of sight (indicated 
with the black dashed lines). A similar phenomenon was observed in another star cluster, 
Blanco\,1 \citep{zhang2020}. This artificial elongation is generated by the distance computed by
simply inverting the parallax, $1/\varpi$. The errors in the parallax measurements $\Delta\varpi$ 
have a symmetric distribution function. After this reciprocation, the distance distribution 
becomes asymmetric. Member candidates in NGC\,2232 and LP\,2439 have $\varpi$ ranging between 
$2.92-3.87$~mas and $2.78-3.46$~mas, respectively. In an earlier study, we performed Monte Carlo 
simulations to estimate the contribution of the error $\Delta \varpi$ on the uncertainty in the 
evaluation of the $X$, $Y$, and $Z$ coordinates \citep{zhang2020}. A typical $\Delta \varpi$ of 
0.19~mas (with a distance in the range $289-360$\,pc), corresponds to an error of $16-25$\,pc for 
our sample.

To correct for the pseudo-elongation generated by parallax errors, we follow the Bayesian 
inversion approach introduced by \citet{bailer2015}. We compute the likelihood based on the measured 
parallax combined with its nominal error with a prior designed for star clusters and field stars. 
The prior assumes a Gaussian spatial distribution of the star cluster members, and an exponentially 
decreasing volume density for field stars \citep{bailer2015}. We adopt the standard deviation of 
cluster-centric distance of individual stars as the scale radius of the Gaussian distribution. For
each star, the field and the cluster prior are combined with weights proportional to the estimated 
membership probability, which we assume 95\%, considering a 5\% field star contamination rate 
(see Section~\ref{sec:stargo}). The corrected distance to each star is the mean value of the posterior 
distribution. A similar method has been applied to the star cluster M67 \citep[][Appendix B]{carrera2019}; 
we refer to the latter work for further details on the method.

\subsection{Interpretation of the 3D morphology}

After having corrected the distance measurements obtained from the Bayesian method, we display 
the projection of member candidates onto the $X-Y$, $X-Z$ and $Y-Z$ planes  (Figure~\ref{fig:3d}, 
panels (d) to (f)). The intrinsic shapes of NGC\,2232 and LP\,2439 are more or less spherical, 
while the purple and green groups resemble filamentary structures stretching out to distances up 
to 50\,pc. Their association with NGC\,2232 is further confirmed by the proximity in the spatial 
distribution. Similar filamentary structures have been found in previous studies. \citet{jerabkova2019} 
discovered the Orion relic filament, which extends 90\,pc in space and is coeval, at an age of 17\,Myr. Additional 250\,pc long filamentary structures with ages in the range of $10-50$\,Myr were identified 
in the Vela~OB2 region by \citet{beccari2020}. When assuming a prompt residual gas expulsion in 
a cluster with a star formation efficiency (SFE) of 1/3, a significantly elongated morphology
will start to appear when star clusters are 80\,Myr old \citep{dinnbier2020b}. Galactic tides need at 
least 100\,Myr to develop two symmetric tidal tails in star clusters \citep{dinnbier2020a}. 
Therefore, the extended structures in both Orion and Vela~OB2 are not tidal tails. This scenario is 
consistent with the observations that open clusters discovered with tidal tails have ages of at least
100\,Myr \citep[see][]{roser2019a, roser2019b, tang2019, zhang2020}. 

According to the numerical study of \citet{dinnbier2020b}, a star cluster with an SFE of 1/3 and 
a phase of rapid gas expulsion (model C10G13), will have about 50\% of unbound members (i.e., outside
the tidal radius) at an age of 25\,Myr. To test this scenario, we compute the tidal radii of NGC\,2232 
and LP\,2439: 
\begin{equation}
    r_t=\left( \frac{GM_C}{2(A-B)^2}\right)^{\frac{1}{3}}\quad ,
\end{equation}
\citep{pinfield1998}. Here, $G$ is the gravitational constant, $M_C$ is the total mass of the star cluster,
and the parameters $A$ and $B$ are the Oort constants 
\citep[$A=15.3\pm0.4\rm~km~s^{-1}~kpc^{-1}$, $B=-11.9\pm0.4\rm~km~s^{-1}~kpc^{-1}$;][]{bovy2017}. 

From the photometric masses of $185.3\pm17.7\rm\,M_\sun$ and $140.5\pm9.4\rm\,M_\sun$, obtained using a 25\,Myr 
isochrone, we compute tidal radii of $8.3\,\pm 0.3\,\rm pc$ and $7.6\,\pm 0.2\,\rm pc$, for LP\,2439 and 
NGC\,2232, respectively. Errors are estimated from the 25\,Myr isochrone with super-solar abundance
(the black dotted curve in Figure~\ref{fig:properties}~(a)). We assume that stars within the tidal 
radius are gravitationally bound to the star cluster, while those beyond the tidal radius are unbound. 
The tidal radius of LP\,2439 is smaller than its half-mass radius of 8.8\,pc, implying that more than 
half of the mass of the stellar grouping is gravitationally unbound. Note that the model cluster (model C10G13) 
in \citep{dinnbier2020b} is ten times more massive than LP\,2439, therefore the required SFE should be higher 
to match the observed bound mass fraction in LP\,2439 (48\%). This implies that the scenario of an SFE higher 
than 1/3 and prompt gas expulsion may provide a good explanation of the observed properties of LP\,2439.

NGC\,2232, on the other hand, has a half-mass radius of 4.9\,pc. Approximately 73\% of its mass is 
gravitationally bound (i.e., within the tidal radius). For both LP\,2439 and NGC\,2232 we observe no
tidal tails beyond the tidal radius. Instead, stars outside the tidal radius are more or less spherically
distributed with respect to the cluster center.

\begin{figure*}[t!]
\centering
\includegraphics[angle=0, width=.95\textwidth]{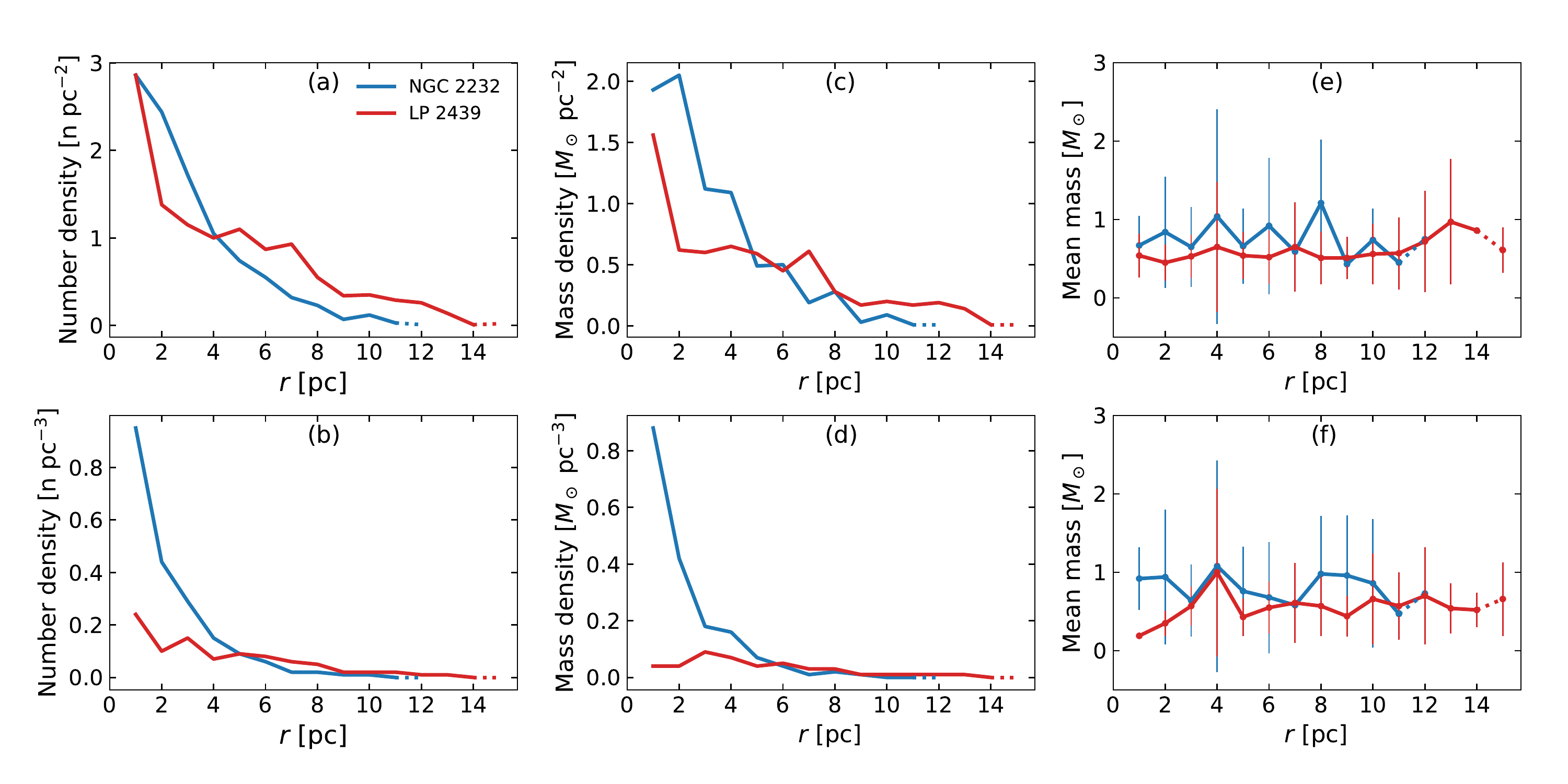}
\caption{
        The number density, mass density, and mean mass distributions along cluster-centric 
            distance $r$ for NGC\,2232 (blue curves) and LP\,2439 (red curves). 
        (a): Surface number density profile; (b): volume number density profile. 
        (c): surface mass density profile; (d): volume mass density profile.
        (e): distribution of mean stellar mass in each surface annulus;
        (f): distribution of mean stellar mass in each volume annulus. 
        The error bar is the standard deviation of the stellar mass in each annulus. The values for the outermost data points (at $r=12$\,pc for NGC\,2232 and at $r=15$\,pc for LP\,2439) 
        are computed using all cluster members with $r>11$\,pc (NGC\,2232) and $r>14$\,pc (LP\,2439).
	    }
\label{fig:annuli}
\end{figure*}

To further quantify the spatial distribution of the stars, 
we plot the radial distribution of the number density, the mass density, and the mean stellar mass
within different annuli of the two star clusters in Figure~\ref{fig:annuli}. The surface number density 
profiles and surface mass density profiles are similar for both clusters, with higher values in the center
and a decline toward larger radius (Figure~\ref{fig:annuli}~(a) and (c)). The volume number density profile and
the volume mass density profile in NGC\,2232 share similar trends as its surface number density profile and surface mass 
density profile (Figure~\ref{fig:annuli}~(b) and (d)). However, the case is quite different in LP\,2439, where 
the volume number density profile and the volume mass density profile are both very shallow (almost flat); see 
Figure~\ref{fig:annuli}~(b) and (d).
The members in LP\,2439 has a larger number  (315) than that of NGC\,2232 (177), and a more extended spatial distribution
(up to 32\,pc) than NGC\,2232 (up to 27\,pc). The difference in surface and volume properties for LP\,2439 
originates in the projection effect of members located in larger 3D annuli onto 2D annuli.

The volume mass density of both NGC\,2232 and LP\,2439 are comparable to those of a ``leaky cluster''
(mass-loss cluster), as
defined in \citet{pfalzner2009}. A mass-loss cluster is undergoing mass loss through 
a range of processes (gas expulsion, stellar 
and dynamical evolution, and tidal field) and expands at a velocity of 2\,pc\,Myr$^{-1}$ at early times. 
The expansion is probably more significant in LP\,2439 since its volume mass density is approaching the 
lower limit of the observed ``leaky clusters'' of 0.07\,$\rm M_\odot~pc^{-3}$ \citep{pfalzner2009}. All 
identified members are brighter than the Gaia DR 2 photometric limit; deeper observations are therefore 
required to probe the contribution of faint members to the density profiles. 

An inspection of the 2D and 3D spatial distribution of the member stars reveals no evidence of mass segregation in
NGC\,2232 and LP\,2439 (see Figure~\ref{fig:annuli}~(e) and (f)). On the other hand, stars enclosed within the $3-4$\,pc 
annulus are generally more massive, and have a somewhat larger dispersion.

\section{Discussion}\label{sec:discussion}

\subsection{Expansions of LP\,2439 and NGC\,2232}\label{sec:expansion}

\begin{figure*}[t!]
\centering
\includegraphics[angle=0, width=.9\textwidth]{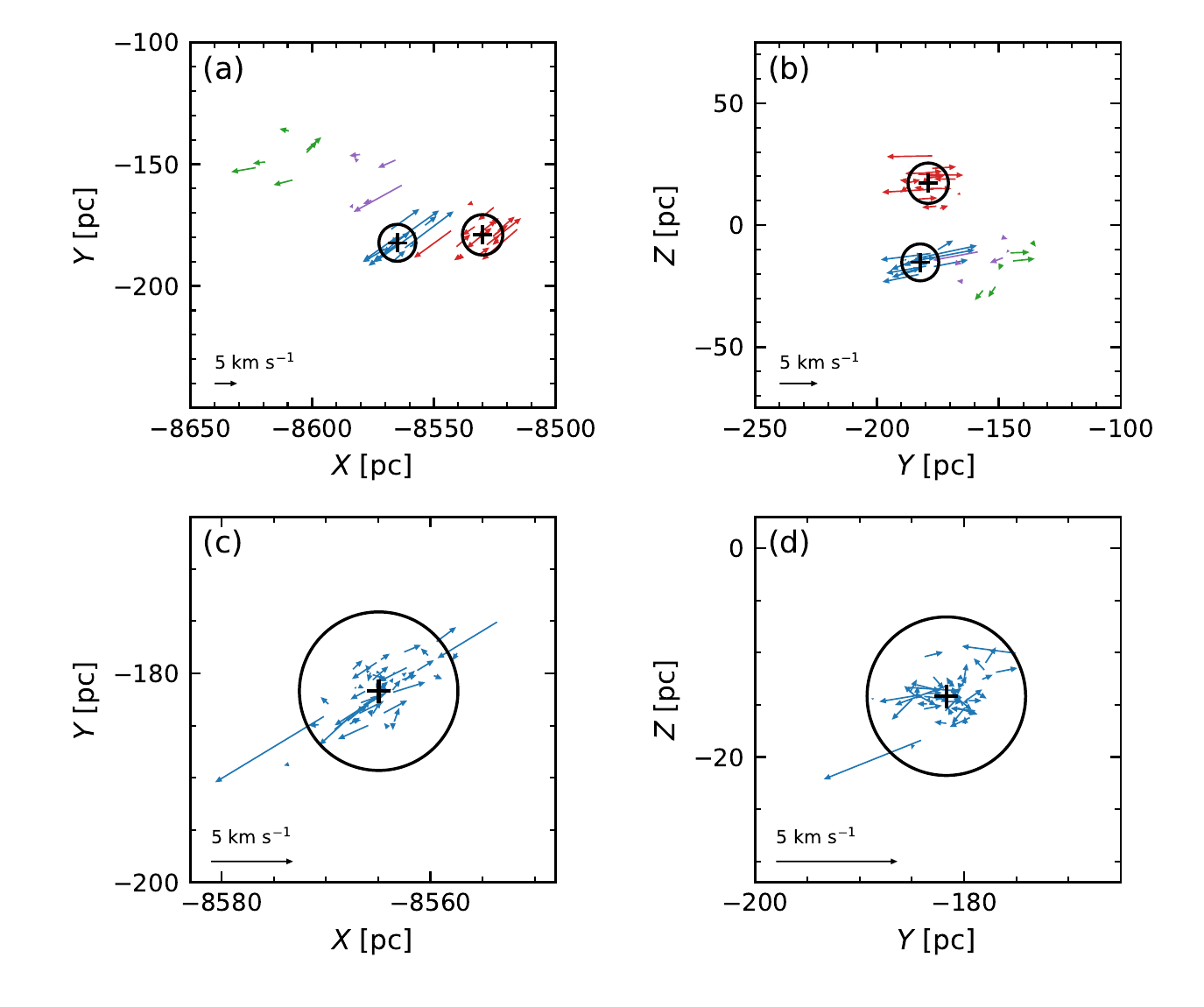}
\caption{
        The relative 3D velocity vectors for member candidates projected onto $X$-$Y$ and $Y$-$Z$ planes.
	    The red vectors represent the velocities of member candidates of LP\,2439 
	        relative to its mean motion, ($U$, $V$, $W$)=($-8.3, 219.1, -4.0$)\,$\rm~km~s^{-1}$, 
	        and the blue vectors the velocities of member candidates of NGC\,2232 relative
	        to its mean motion, ($U$, $V$, $W$)=($-9.3, 219.2, -3.7$)\,$\rm~km~s^{-1}$
	        are shown in panels (a) and (b). RVs in (a) and (b) are obtained from {\it Gaia} DR\,2.
	        The center of each cluster is indicated with the (+) symbol. Black circles denote the tidal radii. The purple
	        and green vectors are the relative velocity of member candidates of the purple and green groups, 
	        relative to the mean motion of NGC\,2232. 
	        (c) and (d): The blue vectors represent the relative velocities of members in NGC\,2232 that 
	        are cross matched with members in \citet{jackson2020} with membership probability larger than 90\%,
	        relative to its mean motion, ($U$, $V$, $W$)=($-9.7, 218.9, -3.8$)\,$\rm~km~s^{-1}$.
	    RVs in (c) and (d) are taken from \citet{jackson2020}. The scale of the velocity vectors is indicated in the bottom-left corner of each panel.	    
	    }
\label{fig:3d_velocity}
\end{figure*}

We investigate the expansion of LP\,2439 and NGC\,2232 using the 3D motions of the member stars
(through combining PMs and RVs). Only 20 members in LP\,2439 have RV measurements, and 17 in 
NGC\,2232. We use the mean positions of the member candidates in each cluster as the respective 
cluster centers of LP\,2439 and NGC\,2232, and the mean velocity in each cluster as the reference 
frame, and present the relative 3D velocity in Figure~\ref{fig:3d_velocity}~(a) and (b). For this
analysis, we consider the purple and green groups associated with NGC\,2232, and use the mean motion 
of NGC\,2232 as the reference velocity for these groups. 

It can be clearly seen that both clusters are expanding, as their members are moving away from the 
cluster center (Figure~\ref{fig:3d_velocity}~(a) and (b)). However, the 3D velocity appears to be 
oriented primarily along the direction of the line-of-sight (see Figure~\ref{fig:3d}). This can be 
explained by the orientation of the velocity error ellipsoid, of which the long axis is aligned with 
the line-of-sight. The typical error  (median value) of {\it Gaia} DR\,2 RV in our sample is 
5.7 $\rm km~s^{-1}$, while the typical error of proper motion is 0.25 $\rm km~s^{-1}$. 
Therefore, the error of the 3D velocity is dominated by the RV error, and the 3D velocity ellipsoid is 
elongated along the line of sight. To reduce this effect, we include RV data of NGC\,2232 from \citet{jackson2020}, 
who observed NGC\,2232 using the Gaia ESO survey with an RV uncertainty of 0.4 $\rm km~s^{-1}$. Among 
the members with membership probability larger than 90\%, 51 members in \citet{jackson2020} (blue vectors) 
cross-match with our members, which demonstrates the expansion within the tidal radius in NGC\,2232 (see Figure~\ref{fig:3d_velocity}~(c) and (d)). A similar expansion is found in the intermediate-age cluster 
Coma Berenices, for stars both within the tidal radius and those located in the tidal tails \citep{tang2019}. 
Since LP\,2439 is a newly-discovered star cluster, no high-resolution spectroscopic observations of its member 
stars are available as of yet.

The cluster expansion is initially spherically symmetric, until the gradient of the Galactic 
gravitational potential deflects the stellar orbits \citep{kroupa2005} and forms tidal tail-like structures. 
The spherical shapes of NGC\,2232, and LP\,2439 suggest that the Galactic tide has just started 
to affect the most distant candidates, and has not yet affected the global morphology of the star clusters.
While majority of stars are moving outward, several members in both NGC\,2232 and
LP\,2439 are falling back to the cluster, which may be a consequence of orbital epicyclic motions 
\citep{boily2003,dinnbier2020a,dinnbier2020b}.


For a 25~Myr-old star cluster to spread over a region of 20~pc in size, an expansion velocity of
at least 0.8\,$\rm km~s^{-1}$ is required. The measured 3D velocity dispersion may be overestimated due to
orbital motion in binary systems \citep[e.g.,][]{kouwenhoven2008}. Since the latter affects RV measurements 
but not PMs, we compute the 2D velocity dispersion from PMs and obtain dispersions of 0.84\,$\rm km~s^{-1}$ 
for LP\,2439 and 0.62\,$\rm km~s^{-1}$ 
for NGC\,2232, with a typical error of 0.25\,$\rm km~s^{-1}$. Considering the above calculation of 
velocity dispersion, the loose spatial distributions in both clusters can be explained well by the 
expansion that started approximately $20-25$~Myr ago.

In addition to gas expulsion, two-body relaxation is another driver of expansion of the systems. 
This dynamical process drives massive stars to gradually sink to the cluster center, and lower-mass 
stars to migrate to the outskirts. However, the radial distribution of the mean stellar mass within 
different annuli does not show evidence  of mass segregation (see Figure~\ref{fig:annuli}~(e) and (f)). 
Therefore, We propose that the expansion in both LP\,2439 and NGC\,2232 is mainly driven by gas expulsion,
which has taken place on a time shorter than the initial timescale of mass segregation.

The gas expulsion effect often induces a gradient of velocity dispersion \citep{kroupa2005}. The 2D 
velocity dispersion obtained from PMs does show a gradually increasing trend from the inner region towards 
the outskirts of the clusters. The 2D velocity dispersion starts at $\sigma_{\rm PM}=0.71~{\rm km~s^{-1}}$ 
within tidal radius, and increases to $\sigma_{\rm PM}=0.94~{\rm km~s^{-1}}$ outside tidal radius for LP\,2439,
and increases from $\sigma_{\rm PM}=0.56~{\rm km~s^{-1}}$ to $\sigma_{\rm PM}=0.72~{\rm km~s^{-1}}$ for NGC\,2232. 
These results are consistent with the predictions of \citet{parmentier2012} and \citet{dinnbier2020a} that an expanding population of stars driven by gas expulsion has a larger velocity dispersion than that of the population within 
tidal radius of the cluster, since escaping stars  generally have ``hotter'' kinematics than the population of gravitationally-bound stars.

\subsection{Different Fates of NGC\,2232 and LP\,2439}\label{sec:fate}

After gas expulsion, NGC\,2232 and LP\,2439 enter a phase of violent relaxation. Whether or not
they can manage to be virialized again depends on the duration of the re-virialization process, 
which typically requires $20-40$ crossing times \citep{parker2016}. To estimate how long it takes 
for cluster to be re-virialized, we compute the crossing time at the moment of gas expulsion, by 
assuming an initial cluster radius of 1\,pc and adopting the 2D velocity dispersion obtained from
proper motion (see Section~\ref{sec:expansion}). 
The crossing times are estimated as $t_{\rm cr} \approx r_h/\sigma$, where $r_h$ is the half-mass 
radius and $\sigma$ is the 2D velocity dispersion obtained from proper motion (see Section~\ref{sec:expansion}). 
The crossing time of NGC\,2232 is 1.6~Myr and that of LP\,2439 is 1.2~Myr. Note that, as the velocity 
dispersion at the onset of gas expulsion is likely higher, the estimated crossing times provide upper 
limits. The age of 25\,Myr corresponds to at least $\sim$17~$t_{\rm cr}$ in NGC\,2232 and 21~$t_{\rm cr}$ 
in LP\,2439.
Considering the 73\% bound mass and the unremarkable density profile (Figure~\ref{fig:annuli}), 
NGC\,2232 may be in the process of re-virialization. On the other hand, with more than half of the 
mass unbound and a shallow volume density profile (Figure~\ref{fig:annuli}), LP\,2439 is most likely 
in the process of dissolution.

Although NGC\,2232 and LP\,2439 originate from the same GMC, we witness two different dynamical 
fates, which stems from the different dynamical states right before the gas dispersal 
\citep{goodwin2009, farias2018}. Assuming the same timescale of gas expulsion and the same SFE in
NGC\,2232 and LP\,2439, we may conclude that NGC\,2232 is initially subvirial and substructured, supported by the 
observed associated filamentary structures. On the contrary, LP\,2439 may be either in a virial or supervirial 
state before the onset of gas expulsion. Since LP\,2439 is more massive, the gas expulsion timescale may be 
much faster in LP\,2439 than in NGC\,2232 considering stronger stellar feedback from massive stars. The 
faster gas expulsion, the faster the expansion of the cluster is. Therefore, this results in a lower survival 
rate of clusters \citep{kroupa2001, baumgardt2007, dinnbier2020b}. 
Either scenario supports the dissolution of LP\,2439. \citet{bravi2018} measured radial velocities of three 
open clusters, IC\,2602, IC\,2391, and NGC\,2547, in the age between 30~Myr and 50~Myr, and find that they 
are also expanding and dispersing. High-resolution spectroscopic data can be used to further constrain the
dynamical history and fate of additional young star clusters in the Galactic field.

\section{Summary}\label{sec:summary}

Utilizing high-precision {\it Gaia} DR\,2 astrometry, we apply the cluster finding method,
\textsc{StarGO}, to identify stellar structures around the NGC\,2232, LP\,2439 and their neighboring
regions in the 5D phase space of stars ($X, Y, Z$, $\mu_\alpha \cos\delta, \mu_\delta$). Additional member 
cleaning is carried out using the location of the candidate members in the color-magnitude diagram. 
LP\,2439 is a cluster newly discovered by \cite{Liu2019} and in this study we quantify for the first
time its dynamical state. Both clusters had their individual distance of member candidates
corrected by a Bayesian method to further discuss their 3D morphology.

\begin{enumerate}

\item We have identified four groupings of stars in the sample: the clusters NGC\,2232 and LP\,2439, 
as well as two neighboring filamentary structures related to NGC\,2232. The identification of the member
candidates of the two star clusters is in a good agreement with earlier studies. The ages of the four 
groups are consistent with an age of 25\,Myr. This suggests that they all formed from the same GMC. 

\item The member stars in LP\,2439 have an average distance of 289.1~pc, with the center at 
R.A.=$103\fdg3785$, Decl.=$-05\fdg8008$, corresponding to Cartesian Galactocentric coordinates of 
($X, Y, Z$) = ($-8526.5$, $-179.1$, $+16.8$) pc. The tidal radius (8.3\,pc) of LP\,2439 is smaller than 
its half-mass radius (8.8\,pc). More than half of the mass in this cluster is gravitationally unbound. 

\item Member stars in NGC\,2232 have an average distance of 323.0~pc, with the center at 
R.A.=$96\fdg8931$, Decl.=$-04\fdg7697$, corresponding to Cartesian Galactocentric coordinates 
($X, Y, Z$) = ($-8564.0$, $-181.3$, $-14.2$) pc.
The tidal radius (7.6\,pc) of NGC\,2232 is larger than its half-mass radius (4.9\,pc). More than 70\%
of the cluster's mass is gravitationally bound. We identify two filamentary structures (which we refer
to as the purple group and the green group) that are associated with NGC\,2232.

\item We observe no evidence of mass segregation among member candidates of LP\,2439 and NGC\,2232. 
We find an overall expansion in both star clusters, that is mainly driven by gas expulsion. The expansion
generates a loose and roughly spherically symmetric spatial distribution of stars. We observe a shallow
volume number density profile and volume mass density profile for LP\,2439.

\item Although NGC\,2232 and LP\,2439 are formed in the same GMC, our analysis suggests that these two 
star clusters will have different dynamical futures. LP\,2439 is undergoing violent dissolution, while NGC\,2232 
may be in the process of re-virialization. We propose that this reflects different initial dynamical 
states prior to gas expulsion in the two clusters. We propose two possibilities to explain why how these star 
clusters have ended up in their current states: 
(i) prior to gas removal, NGC\,2232 may have been substructured and subvirial, while at the same time, 
LP\,2439 may have been either in virial equilibrium or supervirial; or
(ii) gas expulsion in LP\,2439 occurred at a much faster rate than in NGC\,2232.

\end{enumerate}


\acknowledgments
X.Y.P. is grateful to the financial support of two grants of National Natural Science Foundation
of China, No: 11503015 and 11673032. This study is supported by the development fund of Xi'an 
Jiaotong Liverpool University (RDF-18--02--32).  M.B.N.K. expressed gratitude to the National Natural
Science Foundation of China (grant No. 11573004) 
and the Research Development Fund (grant RDF-16--01--16) of Xi'an Jiaotong-Liverpool University (XJTLU).
M.P. acknowledges financial support from the European Unions Horizon 2020 research and innovation program 
under the Marie Sklodowska-Curie grant agreement No. 664931.

We are grateful to the referee for providing in-depth comments and suggestions that greatly help
improving this paper. We send our gratitude to Prof. Dr. Pavel Kroupa and Frantisek Dinnbier for detailed 
discussion on the dynamical evolution of star clusters.
We express thanks to Prof. Dr. Yu Gao and Dr. Hongjun Ma for helpful suggestion on CO emission. 

This work made use of data from the European Space Agency (ESA) mission {\it Gaia} 
(\url{https://www.cosmos.esa.int/gaia}), processed by the {\it Gaia} Data Processing 
and Analysis Consortium 
(DPAC, \url{https://www.cosmos.esa.int/web/gaia/dpac/consortium}). This study also made use of 
the SIMBAD database and the VizieR catalogue access tool, both operated at CDS, Strasbourg, France.


\software{  \texttt{Astropy} \citep{astropy2013,astropy2018}, 
            \texttt{SciPy} \citep{mil11},
            \texttt{TOPCAT} \citep{taylor2005}, and 
            \textsc{StarGO} \citep{yuan2018}
}
{}

\begin{thebibliography}{}

\bibitem[Adams(2000)]{adams2000} Adams, F.~C.\ 2000, \apj, 542, 964

\bibitem[Astropy Collaboration et al.(2013)]{astropy2013} 
            Astropy Collaboration, Robitaille, T.~P., Tollerud, E.~J., et al.\ 2013, \aap, 558, A33

\bibitem[Astropy Collaboration et al.(2018)]{astropy2018} 
            Astropy Collaboration, Price-Whelan, A.~M., Sip{\H o}cz, B.~M., et al.\ 2018, \aj, 156, 123

\bibitem[Bailer-Jones(2015)]{bailer2015} 
            Bailer-Jones, C.~A.~L.\ 2015, \pasp, 127, 994

\bibitem[Banerjee \& Kroupa(2013)]{banerjee2013} Banerjee, S., \& Kroupa, P.\ 2013, \apj, 764, 29

\bibitem[Baumgardt \& Kroupa(2007)]{baumgardt2007} Baumgardt, H., \& Kroupa, P.\ 2007, \mnras, 380, 1589

\bibitem[Beccari et al.(2020)]{beccari2020} Beccari, G., Boffin, H.~M.~J., \& Jerabkova, T.\ 2020, \mnras, 491, 2205

\bibitem[Boily \& Kroupa(2003)]{boily2003} Boily, C.~M., \& Kroupa, P.\ 2003, \mnras, 338, 665

\bibitem[Bovy(2017)]{bovy2017} Bovy, J.\ 2017, \mnras, 468, L63

\bibitem[Bravi et al.(2018)]{bravi2018} Bravi, L., Zari, E., Sacco, G.~G., et al.\ 2018, \aap, 615, A37

\bibitem[Brinkmann et al.(2017)]{brinkmann2017} Brinkmann, N., Banerjee, S., Motwani, B., et al.\ 2017, \aap, 600, A49

\bibitem[Cantat-Gaudin et al.(2018)]{cantat2018} Cantat-Gaudin, T., Jordi, C., Vallenari, A., et al.\ 2018, \aap, 618, A93.

\bibitem[Carrera et al.(2019)]{carrera2019} Carrera, R., Pasquato, M., Vallenari, A., et al.\ 2019, \aap, 627, A119

\bibitem[Claria(1972)]{claria1972} Claria, J.~J.\ 1972, \aap, 19, 303

\bibitem[Cottaar et al.(2012)]{cottaar2012} Cottaar, M., Meyer, M.~R., Andersen, M., et al.\ 2012, \aap, 539, A5

\bibitem[Cropper et al.(2018)]{cropper2018} Cropper, M., Katz, D., Sartoretti, P., et al.\ 2018, \aap, 616, A5

\bibitem[Currie et al.(2008)]{currie2008} Currie, T., Plavchan, P., \& Kenyon, S.~J.\ 2008, \apj, 688, 597

\bibitem[Dame et al.(2001)]{dame2001} Dame, T.~M., Hartmann, D., \& Thaddeus, P.\ 2001, \apj, 547, 792

\bibitem[de Grijs \& Goodwin(2008)]{deGrijs2008} de Grijs, R., \& Goodwin, S.~P.\ 2008, \mnras, 383, 1000

\bibitem[Dinnbier \& Kroupa(2020b)]{dinnbier2020b} Dinnbier, F., \& Kroupa, P.\ 2020, arXiv e-prints, arXiv:2007.00036

\bibitem[Dinnbier \& Kroupa(2020a)]{dinnbier2020a} Dinnbier, F., \& Kroupa, P.\ 2020, arXiv e-prints, arXiv:2006.14087

\bibitem[Farias et al.(2018)]{farias2018} Farias, J.~P., Fellhauer, M., Smith, R., et al.\ 2018, \mnras, 476, 5341

\bibitem[Farias et al.(2015)]{farias2015} Farias, J.~P., Smith, R., Fellhauer, M., et al.\ 2015, \mnras, 450, 2451

\bibitem[Gaia Collaboration et al.(2018a)]{gaia2018a} 
    Gaia Collaboration, Babusiaux, C., van Leeuwen, F., et al.\ 2018, \aap, 616, A10 

\bibitem[Gaia Collaboration et al.(2018b)]{gaia2018b} 
    Gaia Collaboration, Brown, A.~G.~A., Vallenari, A., et al.\ 2018, \aap, 616, A1 
    
\bibitem[Goodwin(2009)]{goodwin2009} Goodwin, S.~P.\ 2009, \apss, 324, 259
    
\bibitem[Goodwin \& Bastian(2006)]{goodwin2006} 
            Goodwin, S.~P., \& Bastian, N.\ 2006, \mnras, 373, 752
            
\bibitem[Hartmann \& Burton(1997)]{hartmann1997} 
            Hartmann, D., \& Burton, W.~B.\ 1997, Cambridge, UK, Cambridge Univ. Press, 243

\bibitem[Hirota et al.(2007)]{hirota2007} Hirota, T., Bushimata, T., Choi, Y.~K., et al.\ 2007, \pasj, 59, 897

\bibitem[Hills(1980)]{hill1980} 
            Hills, J.~G.\ 1980, \apj, 235, 986
    
\bibitem[Jackson et al.(2020)]{jackson2020} Jackson, R.~J., Jeffries, R.~D., Wright, N.~J., et al.\ 2020, \mnras, doi:10.1093/mnras/staa1749

\bibitem[Jerabkova et al.(2019)]{jerabkova2019} 
            Jerabkova, T., Boffin, H.~M.~J., Beccari, G., et al.\ 2019, \mnras, 489, 4418

\bibitem[Juarez et al.(2014)]{juarez2014} Juarez, A.~J., Cargile, P.~A., James, D.~J., \& Stassun, K.~G.\ 2014, \apj, 795, 143 

\bibitem[Kounkel \& Covey(2019)]{kounkel2019} Kounkel, M., \& Covey, K.\ 2019, \aj, 158, 122

\bibitem[Kouwenhoven \& de Grijs(2008)]{kouwenhoven2008} Kouwenhoven, M.~B.~N., \& de Grijs, R.\ 2008, \aap, 480, 103

\bibitem[Kroupa(2005)]{kroupa2005} Kroupa, P.\ 2005, The Three-dimensional Universe with Gaia, 629

\bibitem[Kroupa et al.(2001)]{kroupa2001} Kroupa, P., Aarseth, S., \& Hurley, J.\ 2001, \mnras, 321, 699

\bibitem[Lada \& Lada(2003)]{lada2003} Lada, C.~J., \& Lada, E.~A.\ 2003, \araa, 41, 57 

\bibitem[Levato \& Malaroda(1974)]{levato1974} Levato, H., \& Malaroda, S.\ 1974, \aj, 79, 890

\bibitem[Lindegren et al.(2018)]{lindegren2018} Lindegren, L., Hern{\'a}ndez, J., Bombrun, A., et al.\ 2018, \aap, 616, A2 
    
\bibitem[Liu \& Pang(2019)]{Liu2019} Liu, L., \& Pang, X.\ 2019, \apjs, 245, 32
    
\bibitem[Lynden-Bell(1967)]{lynden1967} Lynden-Bell, D.\ 1967, \mnras, 136, 101

\bibitem[Lyra et al.(2006)]{lyra2006} Lyra, W., Moitinho, A., van der Bliek, N.~S., et al.\ 2006, \aap, 453, 101
    
\bibitem[May et al.(1993)]{may1993} May, J., Bronfman, L., Alvarez, H., et al.\ 1993, \aaps, 99, 105
    
\bibitem[Menten et al.(2007)]{menten2007} Menten, K.~M., Reid, M.~J., Forbrich, J., et al.\ 2007, \aap, 474, 515

\bibitem[Millman et al.(2011)]{mil11} 
        Millman, K. J., Aivazis, M..\ 2011, Computing in Science \& Engineering, 13, 2, 9
    
\bibitem[Moeckel et al.(2012)]{moeckel2012} Moeckel, N., Holland, C., Clarke, C.~J., et al.\ 2012, \mnras, 425, 450

\bibitem[Monroe \& Pilachowski(2010)]{monroe2010} Monroe, T.~R., \& Pilachowski, C.~A.\ 2010, \aj, 140, 2109

\bibitem[Pang et al.(2013)]{pang2013} 
    Pang, X., Grebel, E.~K., Allison, R.~J., et al.\ 2013, \apj, 764, 73 
    
\bibitem[Pang et al.(2011)]{pang2011} Pang, X., Pasquali, A., \& Grebel, E.~K.\ 2011, \aj, 142, 132

\bibitem[Parker \& Wright(2016)]{parker2016} Parker, R.~J., \& Wright, N.~J.\ 2016, \mnras, 457, 3430

\bibitem[Parmentier \& Pfalzner(2013)]{parmentier2013} Parmentier, G., \& Pfalzner, S.\ 2013, \aap, 549, A132

\bibitem[Parmentier \& Baumgardt(2012)]{parmentier2012} Parmentier, G. \& Baumgardt, H.\ 2012, \mnras, 427, 1940

\bibitem[Pfalzner(2009)]{pfalzner2009} Pfalzner, S.\ 2009, \aap, 498, L37

\bibitem[Pinfield et al.(1998)]{pinfield1998} Pinfield, D.~J., Jameson, R.~F., \& Hodgkin, S.~T.\ 1998, \mnras, 299, 955 

\bibitem[R{\"o}ser et al.(2019b)]{roser2019b} R{\"o}ser, S., Schilbach, E., \& Goldman, B.\ 2019, \aap, 621, L2 

\bibitem[R{\"o}ser et al.(2019a)]{roser2019a} R{\"o}ser, S., Schilbach, E., \& Goldman, B.\ 2019, \aap, 621, L2

\bibitem[Rybizki et al.(2018)]{rybizki2018} Rybizki, J., Demleitner, M., Fouesneau, M., et al.\ 2018, \pasp, 130, 74101.

\bibitem[Scandariato et al.(2011)]{Scandariato2001} Scandariato, G., Robberto, M., Pagano, I., et al.\ 2011, \aap, 533, A38

\bibitem[Shukirgaliyev et al.(2017)]{Shukirgaliyev2017} Shukirgaliyev, B., Parmentier, G., Berczik, P., et al.\ 2017, \aap, 605, A119

\bibitem[Tang et al.(2019)]{tang2019} Tang, S.-Y., Pang, X., Yuan, Z., et al.\ 2019, \apj, 877, 12

\bibitem[Taylor(2005)]{taylor2005} Taylor, M.~B.\ 2005, Astronomical Data Analysis Software and Systems XIV, 29

\bibitem[Weiler(2018)]{wei18} Weiler, M.\ 2018, \aap, 617, A138 

\bibitem[Yuan et al.(2018)]{yuan2018} Yuan, Z., Chang, J., Banerjee, P., et al.\ 2018, \apj, 863, 26

\bibitem[Zhang et al.(2020)]{zhang2020} Zhang, Y., Tang, S.-Y., Chen, W.~P., et al.\ 2020, \apj, 889, 99

\bibitem[Zucker et al.(2018)]{zucker2018} Zucker, C., Battersby, C., \& Goodman, A.\ 2018, \apj, 864, 153

\end{thebibliography}
\end{document}